\documentclass[12pt]{article}
\input epsf.sty
\usepackage{hyperref}

\newcommand{\p}{\partial}

\newcommand{\ket}[1]{\left | {#1} \right \rangle}
\newcommand{\bk}[2]{\left \langle {#1} \middle | {#2} \right \rangle}
\newcommand{\bok}[3]{\left \langle {#1} \middle | {#2} \middle | {#3} \right \rangle}

\newcommand{\be}{\begin{equation}}
\newcommand{\ee}{\end{equation}}
\newcommand{\ben}{\begin{eqnarray}\displaystyle}
\newcommand{\een}{\end{eqnarray}}

\usepackage{amssymb}
\usepackage{amssymb}
\usepackage{amsmath}
\usepackage{comment}
\usepackage{mathrsfs}
\usepackage[all]{xy}
\usepackage[dvips]{graphicx}
\usepackage{epsfig,color}
\usepackage[nosort]{cite}

\def\be{\begin{equation}}
\def\ee{\end{equation}}
\def\ba{\begin{align}}
\def\ea{\end{align}}

\def\p{\partial}

\numberwithin{equation}{section}
\begin{document}

\baselineskip=18pt

\begin{center}
{\Large \bf{

On the Stringy Hartle-Hawking State}}

\vspace{10mm}

\vspace{10mm}

\textit{Roy Ben-Israel$\,^1$, Amit Giveon$\,^2$, Nissan Itzhaki$\,^1$ and Lior Liram$\,^1$}
\break

$^1$ Physics Department, Tel-Aviv University, Israel	\\
Ramat-Aviv, 69978, Israel \\

$^2$ Racah Institute of Physics, The Hebrew University \\
Jerusalem, 91904, Israel
\break

\end{center}

\begin{abstract}
We argue that non-perturbative $\alpha'$ stringy effects render the Hartle-Hawking state associated with the $SL(2)/U(1)$ eternal black hole singular at the horizon. We discuss implications of this observation on firewalls in string theory.

\end{abstract}


\section{Introduction}

By now, there are several points of view which suggest that unitarity of the black hole creation and evaporation process leads  to a firewall at the black hole horizon.
't~Hooft's S-matrix approach (see \cite{'tHooft:1996tq} for a review)
seems to require it \cite{Itzhaki:1996jt,Polchinski:2015cea},\footnote{See
however his recent work \cite{Hooft:2015jea}.} the AdS/CFT correspondence \cite{Maldacena:1997re}, in which unitarity  appears to be robust, tends to support this \cite{Marolf:2013dba},  as well as general arguments in quantum information theory \cite{Braunstein:2009my,Mathur:2009hf,Almheiri:2012rt}.

Despite the fact that different approaches seem to support the  firewall scenario, it is hard to see how quantum mechanically the equivalence principle is broken at the horizon.
The main reason  is the lack of a mechanism that could seed the firewall. In other words, forty years later, it is still not clear  what precisely is wrong with the various arguments  from the seventies, which showed that the black hole horizon is smooth  quantum mechanically.

Here, we attempt to challenge the argument due to Israel \cite{Israel:1976ur}. He showed that the  state associated with a scalar field which propagates in the background of eternal black holes is smooth at the horizon, despite the fact that it  is thermal from the point of view of an observer at  infinity. None of the firewall arguments mentioned above is applicable to eternal black holes. Still, we would like to argue that his reasoning, combined with recent non-perturbative input from classical string theory, leads to singularity at the horizon.

Israel's construction is based on the Hartle-Hawking (HH) state  approach \cite{Hartle:1976tp}, in which the Euclidean geometry  sets the initial condition for the Lorentzian system. In the black hole case, the relevant Euclidean geometry is the cigar background. Physics on the cigar geometry
in classical string theory (finite string length scale, $\alpha'=l_s^2$, and parametrically small string coupling, $g_s\to 0$) is different from that in general relativity (GR) in some surprising ways, even for large black holes \cite{Kutasov:2005rr,Giveon:2012kp,Giveon:2013ica,Mertens:2013zya,Giveon:2013hsa,
Giveon:2014hfa,Mertens:2014dia,Mertens:2014saa,Giveon:2015cma,Giribet:2015kca,
Dodelson:2015toa,Mertens:2015hia,Ben-Israel:2015mda}.
This can be shown precisely for the two-dimensional black hole corresponding to the
exact worldsheet CFT, $SL(2, \mathbb{R})_k/U(1)$,
but is believed to be more general.
What is most relevant for us is the fact that the reflection coefficient in the deep UV is different from that in GR in a rather dramatic way -- the density of states associated with it keeps on growing indefinitely \cite{Giveon:2015cma}.  The goal of this paper is to argue that this has far reaching consequences quantum mechanically -- it renders the HH state singular at the horizon.

The paper is organized in the following way. In the next section, we review the Hartle-Hawking state with emphasis on points that are of importance to our discussion. The exact reflection coefficient in $SL(2, \mathbb{R})_k/U(1)$ and its target-space interpretation are discussed in section 3. We observe that modes with different energies disagree about the location of the tip of the cigar. In section 4, we show that this tiny disagreement leads, quantum mechanically, to a rather dramatic effect -- it renders the HH state associated with the black hole singular at the horizon. Section 5 is devoted to attempts to evade this conclusion and,
finally, we summarize with a discussion in section 6.
Some explicit details are presented in the appendix.

\section{The Hartle-Hawking state}

One of the most convincing arguments that the horizon of a quantum black hole is smooth is due to Israel \cite{Israel:1976ur}. He showed  that the HH state associated with eternal black holes is smooth at the horizon, and is thermal from the point of view of an observer at  infinity. Simply put, the black hole  radiation does not render the horizon singular.
In this section, we briefly review the HH state construction with an emphasis on points that will become relevant for the rest of the paper. For a recent review, see e.g. \cite{Harlow:2014yka}.

Suppose that we wish to find the ground state in Minkowski space-time, $\ket{0}_{Min}$,
of some non-trivial theory at $t=0$.
A useful way to do so is to Wick rotate space-time at $t<0$ into Euclidean space, set some arbitrary state $\ket{\Psi}$ at $t=-\infty$, and with the help of the Hamiltonian propagate it to $t=0$. Then, we have
\be\label{lq} \ket{0(t=0)}_{Min}=\lim_{T\to\infty} \exp(-H T )\ket{\Psi(-T)}~.
\ee
Since (\ref{lq}) is correct for any $\ket{\Psi}$, we can write the inner product of $\ket{0(t=0)}_{Min}$ with any state $\ket{\Phi}$ in a path integral form,
\be
\bk{\Phi}{0(t=0)} \sim
\int_{\tilde{\Phi}(t_E=-\infty)=0}^{\tilde{\Phi}(t_E=0)=\Phi} D \tilde{\Phi}\, e^{-I_E}~,
\ee
where $I_E$ is the Euclidean action and $t_E$ is Euclidean time.

\begin{figure}
\centerline{\includegraphics[width=0.9\textwidth,clip=true,trim=70 120 80 120]{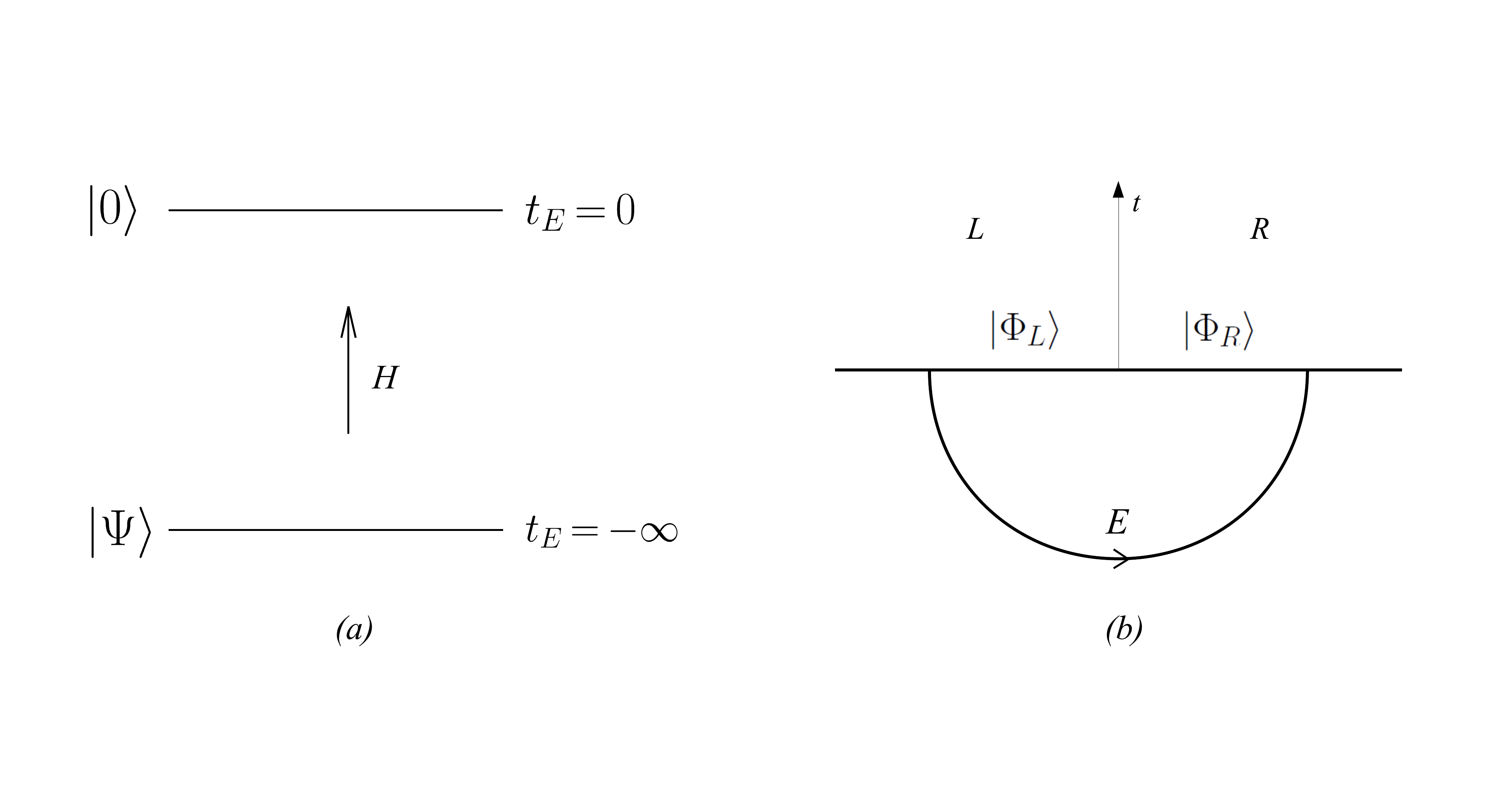}}
\caption{The Euclidean path integral can be evaluated in two ways: (a) by propagating a general state $\ket{\Psi}$ from $t_E=-\infty$ to $t_E=0$, using the Hamiltonian; (b) by propagating $\ket{\Phi_L}$ to $\ket{\Phi_R}$ on the spatial slice $t=0$, using the boost generator.}
\label{pathintegrals}
\end{figure}

This path integral form implies that the same inner product can also be written with the help of the boost generator, which we denote by $E$, that propagates the state on the left (see figure \ref{pathintegrals}) to the state on the right. Schematically,
\be
\bk{\Phi_L \Phi_R }{0(t=0)} \sim \bok{\Phi_L}{e^{-\pi E}}{\Phi_R}~,
\ee
which implies that
\be\label{thermoFieldDouble}
\ket{0(t=0)}_{Min}=\frac{1}{\sqrt{Z}}\,  \sum_{n} e^{-\pi E_n} \ket{E_n}_R \otimes \ket{E_n}_L~,
\ee
where $n$ runs over all states on the left and right.
This way of writing the vacuum, in terms of the eigenstates of $E$, shows  that when tracing over L (R) we get a thermal density state in R (L) with the Rindler temperature $T_{Rindler}=1/2 \pi$.

Being the vacuum state, this state is obviously regular. This regularity, however, is quite fragile in this basis. To see this, we note that (\ref{thermoFieldDouble}) is invariant under boost transformations. Since it is the vacuum, this is consistent with the commutation relation
\be\label{comm}
[E,[E,H]]=H~,
\ee
which requires that any boost invariant state should have either
a vanishing or an infinite Minkowski energy $\langle H\rangle$. The vacuum state (\ref{thermoFieldDouble}) satisfies that with $H\ket{0}_{Min}=0$.

Now, consider a deformation of (\ref{thermoFieldDouble})  that, at first, appears harmless,
\be\label{comma}
\ket{\chi}_{F(E)}=\frac{1}{\sqrt{Z}}\,  \sum_{n} e^{-\pi E_n} F(E) \ket{E_n}_R \otimes \ket{E_n}_L~.
\ee
One would think that if $F(E)$ is a function that differs from $1$ only slightly and/or only at large $E$, then $\ket{\chi}_{F(E)}$ is almost the vacuum, $\ket{0(t=0)}_{Min}$. However, since it is boost invariant and since it is not the vacuum, (\ref{comm}) implies (assuming a unique vacuum) that its Minkowski energy is in fact infinite.

\begin{figure}
\centerline{\includegraphics[width=0.9\textwidth,clip=true,trim=145 180 130 180]{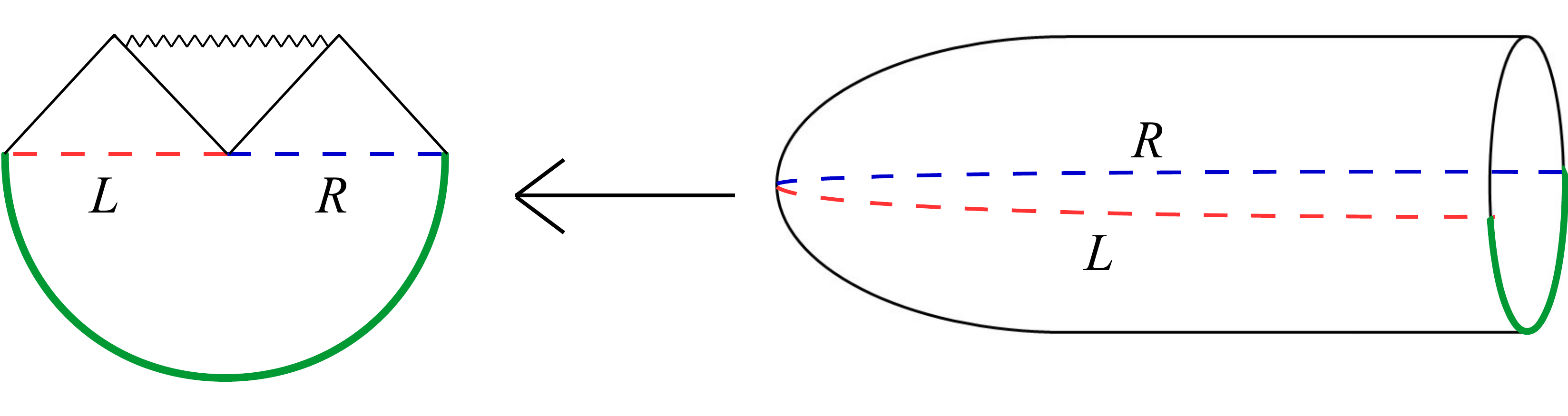}}
\caption{The Hartle-Hawking construction. The vacuum state is constructed by evolving on the cigar and setting the evolved state as the initial condition in the Lorentzian geometry. This can be viewed as gluing half of the cigar geometry to half of the Lorentzian geometry.}
\label{HartleHawkingConstruction}
\end{figure}

For free field theory, $\ket{0(t=0)}_{Min}$ can be written in the form
\be\label{ooa}
\exp \left\{ \int dE  \, e^{-\pi E} \, a^{\dag}_{R,E} \, a^{\dag}_{L,E}
\right\} \ket{0,0}_{Rindler}~,
\ee
where $a^{\dag}_{R(L),E}$ are the creation operators of the Rindler modes on the R (L)
wedge and $\ket{0,0}_{Rindler}$ is the Rindler vacuum. The reasoning above implies that turning on any coupling of these modes will render the state (\ref{ooa}) singular.
This does not mean that there is no vacuum when  interactions are turned on. There is -- it is (\ref{thermoFieldDouble}). Rather, it  means that we should be careful when writing (\ref{thermoFieldDouble}) in terms of $a^{\dag}_{R(L),E}$ in perturbation theory.

We are interested in the HH state associated with the eternal black hole. In this case, we are instructed to glue half of the cigar geometry to half of the eternal black hole background at $t=0$ (see figure \ref{HartleHawkingConstruction}). The boost generator is conserved in the eternal black hole background -- it is the energy as measured by an observer at infinity. Thus, we can use the form (\ref{thermoFieldDouble}) which, after normalization of the energy, reads
\be\label{yh}
\ket{0(t=0)}_{BH}=\frac{1}{\sqrt{Z}}\,  \sum_{n} e^{-\beta E_n /2} \ket{E_n}_R \otimes \ket{E_n}_L~,
\ee
where $\beta$ is the inverse temperature of the black hole. This state is boost invariant, it gives a thermal density when tracing over one side and is smooth at the horizon (that can be approximated by Minkowski space). Hence, from the point of view of an observer at infinity, the black hole radiates with the Hawking radiation, while an infalling observer sees the vacuum.

This concludes the review of the Hartle-Hawking state associated with an eternal
black hole. Next, we shall recall some aspects of the cigar CFT background,
with an emphasis on properties that will have important consequences for the discussion
of the HH state in string theory.

\section{The stringy phase shift on the cigar}

In this section, we review some of the properties of the exact reflection coefficient
in the cigar CFT background.
We start by presenting the setup. Then, we describe the reflection coefficient, \cite{Teschner:1999ug,fzz}, and some of its key features \cite{Giveon:2015cma,Ben-Israel:2015mda}.

\subsection{The setup}

We consider the coset CFT, $SL(2, \mathbb{R})_k/U(1)$, whose sigma-model background takes the form of the cigar geometry
\cite{Elitzur:1991cb,Mandal:1991tz,Witten:1991yr,Dijkgraaf:1991ba},~\footnote{The
eternal black hole is obtained either by Wick rotating the cigar background or,
equivalently, by gauging a spacelike $U(1)$ of the underlying $SL(2,\mathbb{R})$
model instead of the timelike one.}
\be\label{cigar}
ds^2=2k \tanh^2 \left(\frac{r}{\sqrt{2 k}}\right) d\theta^2
+dr^2~,~~~~
\exp(2 \Phi )= \frac{g_{0}^2}{\cosh^2 \left(\frac{r}{\sqrt{2 k}}\right)}~;
\ee
the angular direction $\theta$ has periodicity $2\pi$, compatible with
smoothness of the background at the tip, and $\Phi$ is the dilaton.
We work with $\alpha'=2$.
In the supersymmetric case,
the background (\ref{cigar}), which is obtained e.g.
by solving the graviton-dilaton e.o.m. in the leading GR approximation,
is perturbatively {\it exact} in $\alpha'$ \cite{Bars:1992sr,Tseytlin:1993my}.

\begin{figure}
\centerline{\includegraphics[scale=0.35]{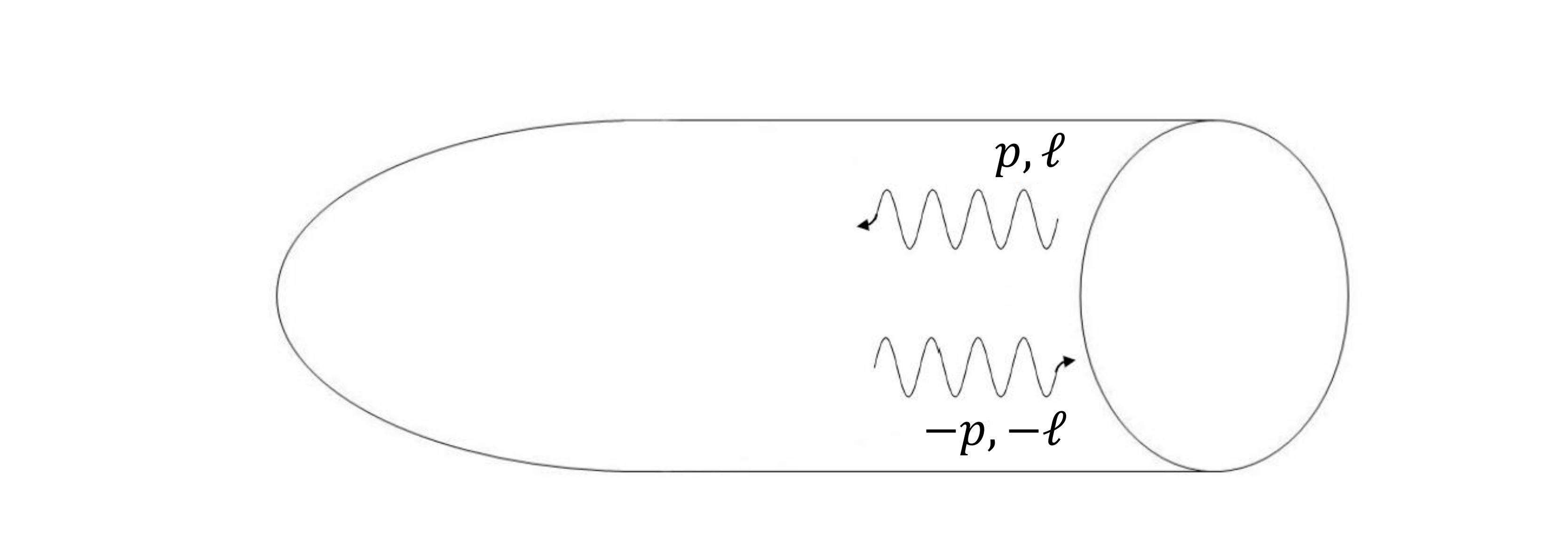}}
\caption{The classical geometry induced from the $SL(2, \mathbb{R})_k/U(1)$ coset model -- a cigar that pinches off at $r=0$, which upon Wick rotation, corresponds to the horizon. Asymptotically, the classical solution of the Klein-Gordon equation is a sum of incoming and outgoing radial waves with radial momentum $p$ and angular momentum $\ell$.}
\label{cigarplot}
\end{figure}

The reflection coefficients associated with modes that scatter in this background (see figure \ref{cigarplot}) are known exactly in $\alpha'$ \cite{Teschner:1999ug} and, as we shall review, have some fascinating features.
In terms of the underlying $SL(2,\mathbb{R})$,
these modes are in the continuous representations, namely, with $SL(2,\mathbb{R})$
quantum numbers $(j;m,\bar m)$ given by
\be\label{js}
j=-1/2+is, ~~s\in \mathbb{R}~,\qquad (m,\bar m)={1\over 2}(\ell+kw,-\ell+kw)~,
~~\ell,w\in \mathbb{Z}~,
\ee
where $\ell$ is the angular momentum along the $\theta$ direction,
$w$ is the winding on the cigar, which we will set to zero here,
and $s$ is related to the momentum in the radial direction;
more precisely,
\be
s=\sqrt{\frac{k}{2}} \, p~,
\ee
where $p$ is the momentum associated with the canonically-normalized field
corresponding to the radial direction at $r\to\infty$.

\subsection{The exact reflection coefficient}

The reflection coefficient in the exact $SL(2,\mathbb{R})/U(1)$ CFT is obtained
by inspecting the asymptotic behavior of vertex operators corresponding,
in our present case of interest,
to $j=-1/2+is$, $m=-\bar m=\ell/2$ (see e.g. \cite{Aharony:2004xn}),
\be\label{vvv}
V_{j;m,\bar m}\sim\left(e^{ipr}+R(p,\ell) \, e^{-ipr}\right)~.
\ee
The two point function of such operators, normalized such that the coefficient of their incoming wave is 1, gives the exact reflection coefficient;
it takes the form
\be\label{ftg}
R(p,\ell)= R_{pert}(p,\ell)~ R_{non-pert}(p)~.
\ee
This was originally obtained via the bootstrap approach \cite{Teschner:1999ug,fzz} and was verified using the screening operators approach in \cite{Giribet:2000fy,Giribet:2001ft}.

The perturbative part, $R_{pert}$,
can be derived from GR plus perturbative $\alpha'$ corrections.
In the supersymmetric case, on which we focus here,
there are no perturbative $\alpha'$ corrections, and
$R_{pert}$ can thus be derived by solving the wave equation in
the curved background (\ref{cigar}). For momentum modes ($w=0$), this gives
\be
R_{pert}=e^{i\delta_{pert}}=\nu^{2i s} \frac{\Gamma(2i s)}{\Gamma(-2is)} \, \frac{\Gamma^2 \left( \frac12 + \frac{\ell}{2} -i s \right)}{\Gamma^2 \left( \frac12 + \frac{\ell}{2} +i s \right)}~, \label{Rpert}
\ee
where $\nu$ is a parameter which is related to
the location of the tip of the cigar
(see e.g.~\cite{Aharony:2004xn} and references therein).

The non-perturbative correction in $\alpha'$, $R_{non-pert}$,
is due to the condensate of a winding string mode near the tip of the cigar
(the Sine-Liouvile potential) \cite{fzz,kkk}; it reads
\be
R_{non-pert}=e^{i\delta_{non-pert}}=\frac{\Gamma(1+\frac{2is}{k})}{\Gamma(1-\frac{2is}{k})}~. \label{stringyphase}
\ee
As expected, $R_{non-pert}$ is controlled by a scale that is much shorter than the  $R_{pert}$ scale (for large $k$).

We note (\ref{ftg}) has three distinct features that are key for its target-space interpretation:
\begin{itemize}
\item
It is a product of two functions. Namely, the non-perturbative phase shift, $\delta_{non-pert}$, is simply added to the perturbative phase shift, $\delta_{pert}$. Normally, in quantum mechanics, when a correction is added to the potential, its effect on the exact phase shift is highly non-trivial. A factorization like in (\ref{ftg}) calls for an explanation.
\item
While $R_{pert}$ depends both on the radial and angular momenta, $R_{non-pert}$ depends only on $p$. This is also quite peculiar. Even if there is a factorization, one would expect $R_{non-pert}$ to depend on $\ell$ as well.
\item
In the deep UV (large $p$), $R_{non-pert}$ dominates $R_{pert}$ \cite{Giveon:2015cma}. To be more precise, as dictated by general reasoning, the density of states associated with  $R_{pert}$ goes to a constant in the UV,
\be\label{ft}
\rho_{pert}(p)\equiv \frac{1}{2\pi i} \frac{d \left(\log R_{pert}\right)}{dp} = \sqrt{\frac{k}{2}} \, \frac{\log(4 \nu)}{\pi} + \frac{1-2\ell^2}{4\pi \sqrt{2k}}\,\frac{1}{p^2} + \cdots~.
\ee
On the other hand, the density of states associated with $R_{non-pert}$ grows logarithmically with $p$,
\be
\rho_{non-pert}(p)\equiv \frac{1}{2\pi i} \frac{d \left(\log R_{non-pert}\right)}{dp} = \sqrt{\frac{2}{k}} \, \frac{\log \left(p \right)}{\pi} + \cdots~,
\ee
for $p\gg \sqrt{k}$.
\end{itemize}

Let us see what is the natural target-space interpretation of these features. Perturbative corrections are expected to stay in the same ``universality class" of (\ref{ft}). Namely, they could modify $\nu$ and  the sub-leading terms in (\ref{ft}), but cannot induce more dominant terms. For example,  in the bosonic case there are corrections to the reflection coefficient which amount to a shift $k\to k-2$. As is clear from (\ref{ft}), such a shift can be absorbed into a redefinition of $\nu$ plus terms that vanish in the UV.

\begin{figure}
\centerline{\includegraphics[scale=0.30]{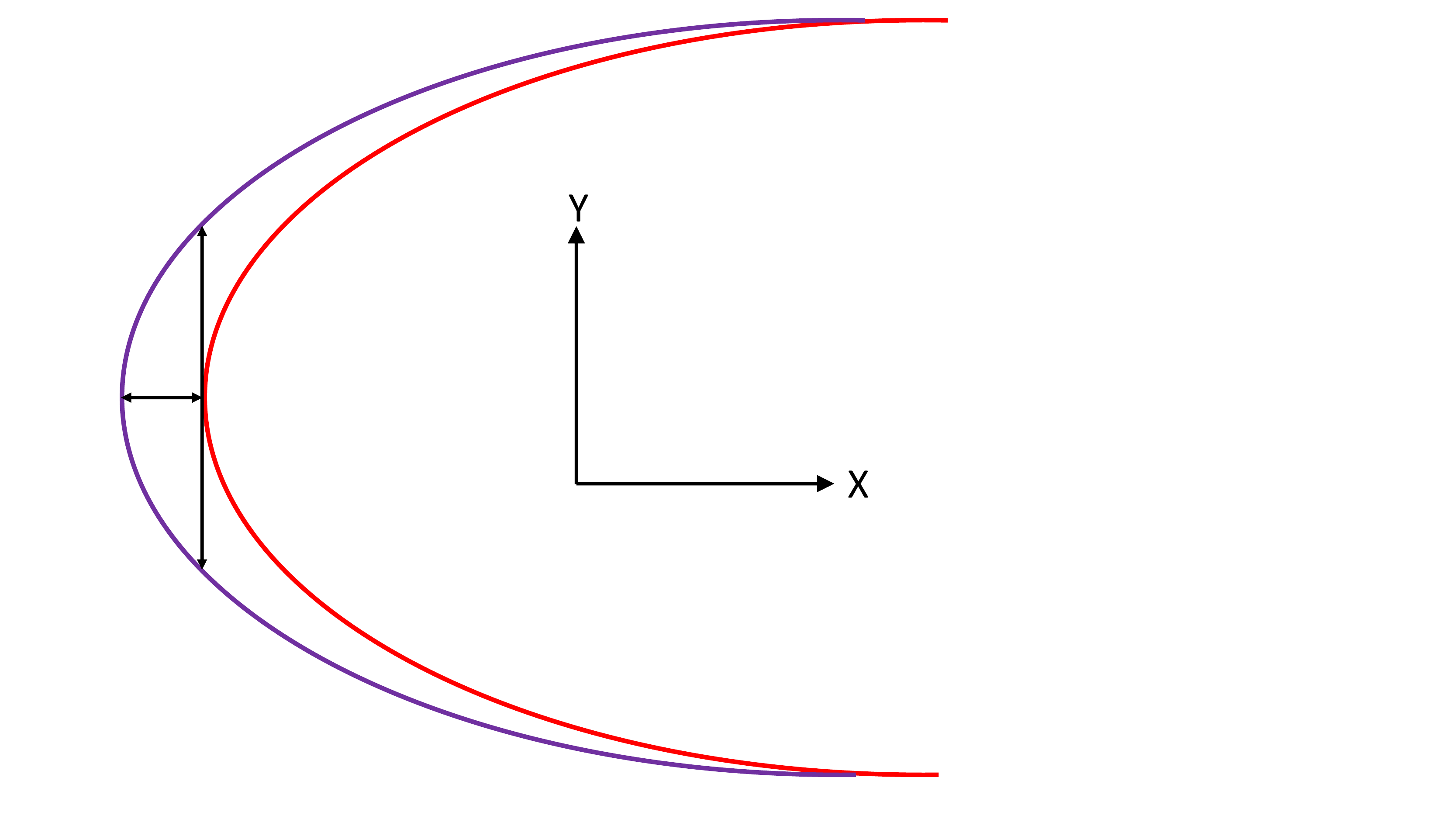}}
\caption{The cigar appears longer for more energetic modes. For large $k$, typically the effect is tiny, $\Delta X\sim 1/\sqrt{k}$. However, since the radius of curvature of the cigar scales like $\sqrt{k}$, we see that at the tip of the cigar $\Delta Y \sim 1$. A different reasoning, which is based on the FZZ duality \cite{fzz,kkk}, led to the same conclusion \cite{Giveon:2015cma}. }
\label{fig1}
\end{figure}

The fact that  $\rho_{non-pert}(p)$ blows up in the UV implies that the non-perturbative effects do not fall into the same universality class as (\ref{ft}): they cannot be accounted for by normalization of $\nu$ and a smooth change in the shape of the cigar.  The reason is that normalizing $\nu$ ``adds" a finite amount of space to the geometry (\ref{ft}), while $\rho_{non-pert}(p)$ is unbounded in the deep UV.
A natural target-space description of the non-perturbative effect is the following: As we go deeper and deeper into the UV, the cigar becomes longer and longer. Its shape is left unchanged -- it is just that the location of the tip depends on $p$ (see figure \ref{fig1}).
The extra distance the modes have to propagate, $\Delta X(p)$ in figure \ref{fig1}, is the cause of $\rho_{non-pert}$.
This explains both the factorization and the lack of dependence of $R_{non-pert}$ on the angular momentum. The cause of $R_{pert}$ is the shape of the cigar; since it remains the same as we go deeper into the UV, there are no corrections to $R_{pert}$.
On the other hand, $R_{non-pert}$ is due to the extra distance the modes have to propagate, $\Delta X(p)$, that clearly does not depend on the angular momentum.

\section{The Stringy HH state}

Next we turn to the implications of the point of view presented in the previous section to Israel's conclusion.

Neglecting interactions, the HH state takes the form
\be\label{hhw}
\ket{HH} = \exp \left\{ \int dE  \, e^{-\beta E/2} \, a^{\dag}_{R,E} \, a^{\dag}_{L,E}
\right\} \ket{0,0}~,
\ee
where $a^{\dag}_{R(L),E}$ are the creation operators of canonically normalized modes with (Schwarzschild) energy $E$ on the R (L) side, and $\ket{0,0}$ is the analog of the
Rindler vacuum for the eternal black hole. As discussed in section 2, this state  has the following properties:
\begin{itemize}
\item
If we trace over L (R), we get a thermal density state on R (L).
\item
It is invariant under the boost symmetry of the eternal black hole.
\item
It is smooth at the horizon.
\end{itemize}

\begin{figure}
\centerline{\includegraphics[scale=0.45]{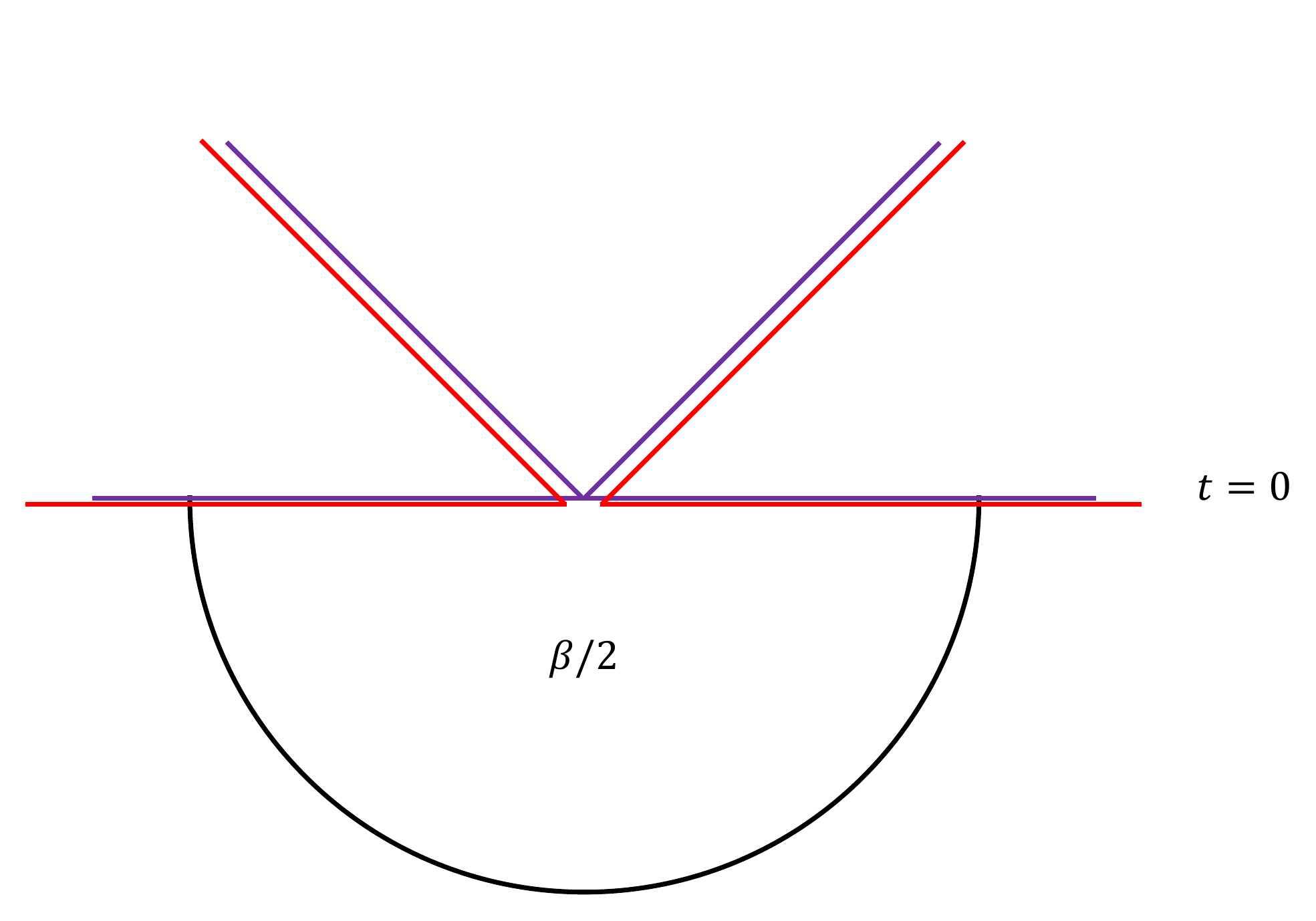}}
\caption{The usual HH construction requires gluing of the Euclidean geometry (evolved over half of the compact time) to the Lorentzian geometry at the surface $t=0$, generating a state that is regular across the purple surface. But with the stringy corrections, high-energy modes see a longer cigar and so there is a mismatch in the gluing, namely, in the two sides of the black hole (depicted in red are low-energy modes, which in comparison travel a shorter distance than the purple ones). This mismatch cannot be absorbed into the metric because of the non-perturbative nature of the correction (the slight seperation
of the two energies on the $t=0$ line was made for presentation purposes only).}
\label{fig5}
\end{figure}

What happens in string theory?
In string theory, the creation operators that appear in (\ref{hhw}) are the ones that appear in the S-matrix. Hence, on the cigar geometry, these are the modes that lead to (\ref{ftg}). We saw in the previous section that these modes disagree about the location of the tip. Thus, the point where the Euclidean tip is glued with the Lorentian horizon  depends on the energy (see figure \ref{fig5}). This looks like a tiny and negligible effect. However, since it simply does not occur in GR (including perturbative $\alpha'$
corrections), this is a completely novel non-perturbative stringy effect in $\alpha'$. Hence, it is hard to be certain that its effect is negligible. In fact, we next argue it has far reaching consequences quantum mechanically.

There are two ways to address the implications of this mismatch at the tip. One is to work with the original modes and take account of the mismatch in the HH state. An equivalent way, that we follow below,  is to define new modes that agree on the location of the tip, write down the  HH state in terms of these modes and see what we get.

The creation operators of modes that agree on the location of the tip, which we denote by $b^{\dag}(p)$, are related to the creation operators of natural stringy modes,
$a^{\dag}(p)$, that appear in string scattering amplitudes,
by stripping off the stringy phase,
\be
b^{\dag}(p)=e^{-\frac{i}{2}\delta_{non-pert}}a^{\dag}(p)~.
\ee
This is obtained, e.g., via the shift of $r$ in (\ref{vvv})
required to get rid of the $e^{i\delta_{non-pert}}$ phase in $R(p,\ell)$,
and consequently match the red and purple lines in figures \ref{fig1} and \ref{fig5}.
Written in terms of these modes, the HH state (\ref{hhw}) takes the form
\be\label{thh}
\ket{HH} = \exp \left\{ \int dE \, e^{-\beta E/2} \, e^{i \theta(E)} \, b^{\dag}_{R,E}  \, b^{\dag}_{L,E} \right\} \ket{0,0}~,
~~~\mbox{with }~~~
e^{i \theta(E)}=e^{i\delta_{non-pert}}~,
\ee
where here, for simplicity, we focus on massless modes and use the on-shell condition, $E=p$.

Since $b^{\dag}(p)$ create modes that agree on the location of the tip, we can study the properties of (\ref{thh}) as if it was a state propagating in a GR background
(or in a string theory background including perturbative $\alpha'$ corrections), in which the location of the tip does not depend on the energy. Clearly, (\ref{thh}) is boost invariant  and it gives a thermal density matrix upon tracing over one side. However, the discussion around  (\ref{comma}) implies that, for any non-trivial $\theta(E)$, the state (\ref{thh}) is singular -- it has an infinite amount of energy  as measured by an infalling observer.\footnote{It is worth emphasising that none of this happens in  Minkowski space-time; there, the thermofield double state in string theory is the usual one, without an extra phase.}
Indeed, a calculation of the  energy as observed by an infalling observer gives (see the appendix for details)
\begin{equation}\label{tote}
\langle H \rangle  =  \frac{\sqrt{2k}}{\pi }\int_{0}^{\infty} dE\, \frac{\sin^2 \left(\frac{\theta(E)}{2} \right)}{\sinh^2 \left( \pi\sqrt{2k} E \right)}  \, \int_{0}^{\infty} dP~,
\end{equation}
where $P$ is the Kruskal energy and, as before, $E$ is the Schwarzschild energy. Note that, indeed, for any non-trivial $\theta(E)$, this expression is divergent due to the integral over $P$, and it cannot be regularized without violating boost invariance. In section 5,
we elaborate further on (\ref{tote}).

Boost invariance also implies that this infinite energy must be concentrated on the horizon. A simple way to see this is to recall that in two dimensions free massless fields are decomposed into left and right moving modes. The only left (or right) moving modes that are boost invariant are localized on the horizon. Again, one can verify this with a direct calculation (the details are in the appendix) by point-splitting the energy-momentum tensor. For points on the same side of the horizon ($V_1 V_2 >0$ for left moving modes;
see figure \ref{Vcoord})
we get the standard answer, which does not depend on $\theta(E)$.
However, with legs on both sides of the horizon ($V_1 V_2 <0$), the correlator (of e.g. a free massless scalar field) does depend on $\theta(E)$,
\begin{eqnarray}
\langle \partial \phi (V_1) \, \partial \phi (V_2) \rangle &=& \frac{k}{2\pi V_1 V_2} \int_0 ^\infty dE  \, \frac{E \cos \left( \theta(E) + E \tau \right)}{\sinh \left(\pi \sqrt{2k} E \right)}~, \label{FNS_1}
\end{eqnarray}
with $\tau \equiv \sqrt{2k} \log(-V_1/V_2)$. Therefore, the only contribution to (\ref{tote}) comes from the horizon itself (see figure \ref{Vcoord}).

\begin{figure}
\centerline{\includegraphics[width=0.30\textwidth,clip=true,trim= 180 300 160 280]{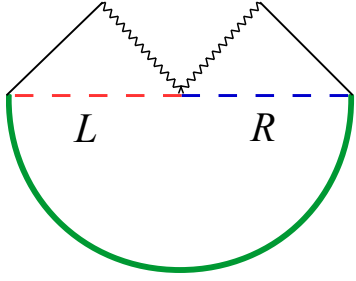}}
\caption{The non-trivial stringy phase renders the Hartle-Hawking state  singular at the horizon.}
\label{singularHorizon}
\end{figure}

To summarize, the picture we seem to encounter is the following: The flux measured by an observer at infinity is the same as the usual thermal flux with the same Hawking temperature; however, on the horizon there is an infinite flux that is seen only by an infalling observer and not by the observer at infinity. Thus, an infalling observer will encounter a naked null singularity at the horizon, rendering the two sides of the eternal black hole geometry being practically disjoint (see figure \ref{singularHorizon}).

\section{Evading the singularity at the horizon?}

In the previous section we presented some compelling arguments that if we follow the HH construction of a black hole vacuum in string theory -- at least for strings propagating in the $SL(2, \mathbb{R})_k/U(1)$ CFT background -- quantum mechanically, the horizon is singular. Here, we consider possible ways to evade this conclusion, and in the process we expose the nature of the singular horizon.

\subsection{Can we slice away the singularity?}
\begin{figure}
\centerline{\includegraphics[scale=0.5]{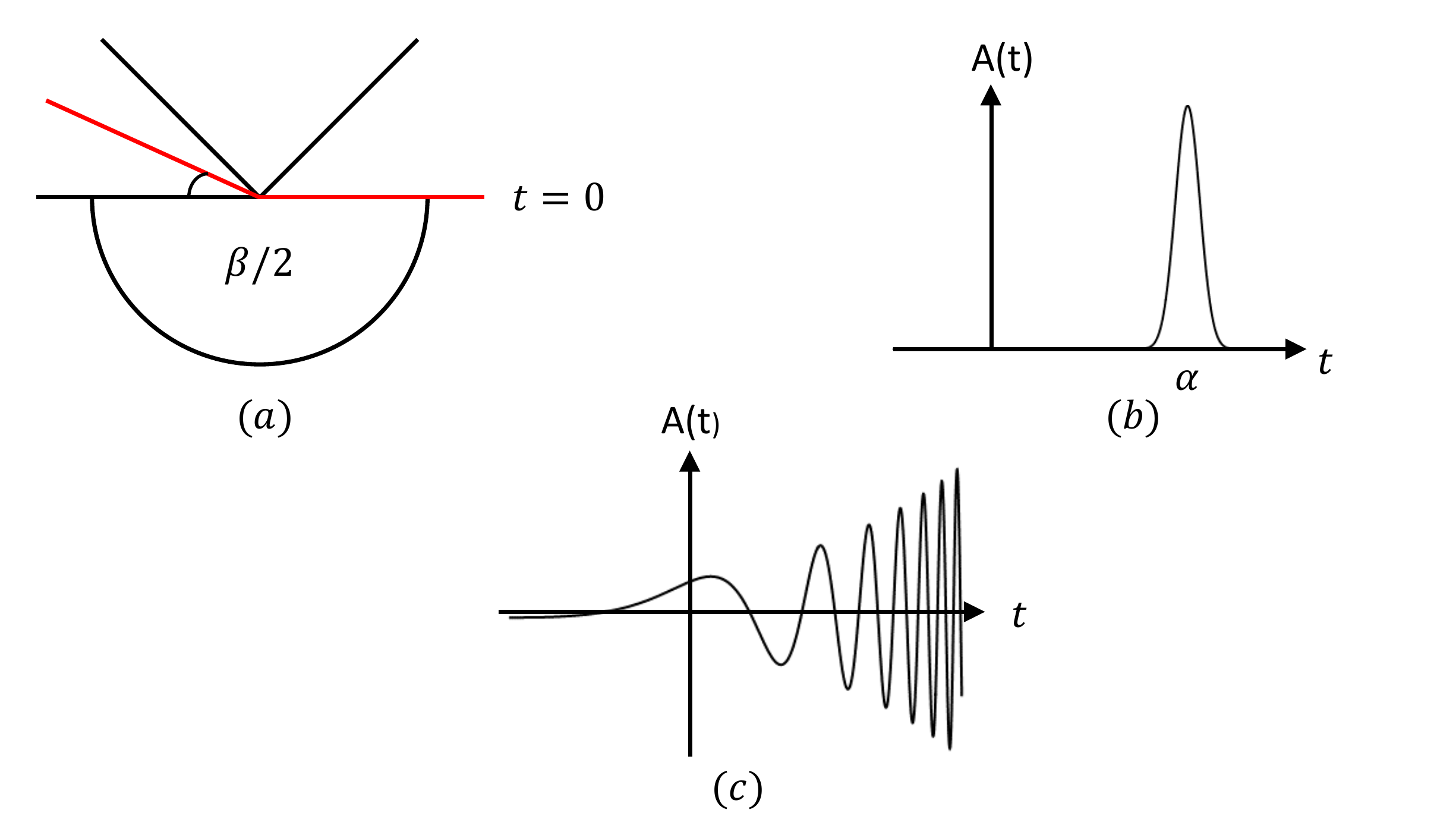}}
\caption{The relative amplitude of the wave function on each slice of constant Rindler time $t$. Figure (a) shows that for the case (\ref{phaselinear}), the wave function has support only on a constant time slice $t=\alpha$, and so the quantization surface is classically redefined (with no uncertainty). Figure (b) shows the amplitude for the case (\ref{wq}), where the amplitude is very large around $t=\alpha$ and small otherwise, and figure (c) shows the amplitude for the $\delta_{non-pert}$ stringy case. In the stringy case, the amplitude is not localized with finite uncertainty and diverges as $t \to \infty$, namely, as we approach the horizon.}
\label{amplitudes}
\end{figure}

We argued that (\ref{thh}) is singular at the horizon with the standard slicing at $t=0$. The question is whether there is a different slicing that renders (\ref{thh}) regular. The fact that this is a valid possibility can be illustrated with the help of the following example \cite{Maldacena:2013xja}. Suppose that
\be\label{phaselinear}
\theta(E)=\alpha E~,
\ee
with some constant $\alpha$.
Such a $\theta(E)$, being non-trivial, leads to a singularity at the horizon. However, this singularity is a fake one. A simple way to see this is to note that in the Euclidean background, the phase (\ref{phaselinear}) can be absorbed into a redefinition of $\nu$ in the cigar background (see (\ref{ft})). Indeed, in the HH state we can slice space-time in such a way that exactly cancels this phase \cite{Maldacena:2013xja}.
The reason why the phase is cancelled is that instead of rotating L to R with $E$ an angle $\beta/2$, we do so for $\beta/2 + i \alpha$ (see figure \ref{amplitudes}(a)). The extra piece, $i\alpha$, is the contribution of $\theta$ in this case.

A smooth non-linear $\theta(E)$ whose non-linearity is controlled by the curvature of the black hole and whose asymptotic behaviour is
\be\label{wq}
\theta(E) \xrightarrow{E \to \infty} \alpha E~,
\ee
also leads to a fake singularity. In the Euclidean setup, this can be absorbed into a redefinition of $\nu$ combined with a smooth change in the shape of the cigar.
In string theory, however, $\theta(E)$ grows faster than (\ref{wq}) at large $E$,
\be\label{uvir}
\theta_{stringy}(E)=\delta_{non-pert}(E)\sim E \log(E)~.
\ee
As discussed above,
such a phase cannot be accounted for by rescaling $\nu$
and modifying smoothly the shape of the cigar.
Hence, there is no suitable way of slicing Lorentzian space-time
and glue it to a half cigar shape geometry,
as required to construct a regular HH-like state.

We can also study these three typical cases from the viewpoint of the wave function, which will provide further insight into fundamental difference between them.
As evident from (\ref{thh}), the HH state is a superposition of modes with definite energy, with a relative phase between them. A Fourier transform gives a superposition of wave functions at different times with different relative amplitudes -- each mode would be boosted in Lorentzian space by an angle dependent on its energy. For the first case, (\ref{phaselinear}), the Fourier transform of the relative phase\footnote{When Fourier transforming the relative coefficients of the modes, we ignore the factor of $e^{-\beta E/2}$ since it just means that in Euclidean time we evolve exactly on the half-circle.} trivially gives $\delta(t-\alpha)$, assuring that the wave function is defined exactly on that particular time slice and vanishes everywhere else. 

In the second case, (\ref{wq}), the picture is somewhat different. The Fourier transform does not give a delta function, so the wave function does not vanish outside the slice $t=\alpha$, but the relative amplitude of this surface is much larger than any other surface (this can be seen through the saddle point approximation, or simply by noting that for $t=\alpha$, the integrand of the Fourier transform tends to 1). While in the previous case our wave function had support only on the surface $t=\alpha$, and could be thought of as a classical slicing (in the sense that the initial condition wave function is exactly localized in time), in the asymptotically linear case we have a semi-classical slicing; the initial condition is localized on a surface with a small amount of uncertainty, thus creating small corrections around the classical slicing.

In the third case, where $\theta(E) = \delta_{non-pert}(E)$, the behavior of the Fourier transform of the phase is very violent (see figure \ref{amplitudes}). For large $t$, it is given by (see \cite{Natsuume:1994sp,Ben-Israel:2015mda})
\begin{equation}
\int_{-\infty}^{\infty} dE \, e^{i E t} \,
\frac{\Gamma\left(1+i\sqrt{\frac{2}{k}}E\right)}{\Gamma\left(1-i\sqrt{\frac{2}{k}}E\right)} \approx  -\sqrt{\frac{k}{2\pi}} \,  e^{\frac14 \sqrt{\frac{k}{2}} \, t} \cos \left(2  e^{\frac12 \sqrt{\frac{k}{2}} \, t}   +\frac{\pi}{4}\right). \label{2ptFTstresasymp}
\end{equation}
This means not only that the wave function is not at all localized in time, but also that the amplitude blows up as it approaches the horizon (at $t \to \infty$). This could be viewed as a highly quantum slicing in which it is impossible to define an initial condition for the black hole on a spacelike slice with finite uncertainty. 
We note that this is a direct consequence of the UV/IR mixing discussed in \cite{Ben-Israel:2015mda}, entered through the attempt to glue the Euclidean geometry to the Lorentzian one on a constant slice.

Finally, let us return to (\ref{tote}), to better understand the origin of the divergent energy. Seemingly, most of the contribution comes from low-energy modes (because of the exponential suppression of the integrand). But in fact, we see that these modes, whose behavior is approximately linear in $E$, only redefine the slicing and the divergence due to them is not real. The only real divergence comes from high energies which, despite giving a small result in the first integral of (\ref{tote}), are multiplied by a divergent quantity (the second integral in (\ref{tote})) and cannot be fixed by absorbing into $\nu$ and smoothly deforming the shape of the cigar. 
This illustrates again the importance of the UV/IR mixing in  black hole physics.

\subsection{Can we twist away the singularity?}
We claimed that the HH state is singular at the horizon and that this singularity cannot be removed by a suitable slicing of space-time (as it is spread all over the time axis).
Is it possible, though, that a different state is regular at the horizon while being boost invariant and yielding a thermal spectrum at infinity?

At first, the answer seems to be yes. Consider the following unitary ``twist" of the HH state,
\be\label{ty}
\ket{\widetilde{HH}}=U(\theta)\ket{HH}= \exp\left\{ \int dE \,  e^{-\beta E/2} \, e^{-i \theta(E)} \, a^{\dag}_{R,E} \, a^{\dag}_{L,E} \right\}  \ket{0,0}~.
\ee
Just like the standard HH state, this state is boost invariant and it leads to a thermal density when tracing over one side. When written in terms of the modes that agree on the location of the tip/horizon, we have
\be\label{po}
\ket{\widetilde{HH}} = \exp \left\{ \int dE \, e^{-\beta E/2}  \, b^{\dag}_{R,E} \, b^{\dag}_{L,E} \right\} \ket{0,0}~.
\ee
This looks just like (\ref{ooa}) (with $\beta$ instead of $2\pi$), and so it appears to be regular at the horizon.

There is, however, an important difference between (\ref{ooa}) in a GR background and (\ref{po}) in string theory. In a GR geometry, (\ref{ooa}) is the limit of (\ref{yh}) at zero coupling. As discussed in section 2, at finite coupling, (\ref{ooa}) becomes singular and the regular state (that is thermal for an observer at infinity and is boost invariant) is (\ref{yh}). Similarly, in string theory, (\ref{po}) is regular at zero coupling and it becomes singular at any finite coupling. However, (\ref{po}) is not the zero coupling limit of  (\ref{yh}). The zero coupling limit of (\ref{yh}) in string theory is (\ref{thh}), which is singular at the horizon. We do not know how to twist, for $\theta(E)$ whose UV divergence is faster than $\alpha E$, the state (\ref{yh}), in such a way that its zero coupling limit gives (\ref{po}).

This is closely related to the fact that from the point of view of string theory the natural modes are the ones that are created by $a^{\dag}(p)$. The modes that are created by $b^{\dag}(p)$ interact in a highly non-local fashion.
As recalled above, (\ref{uvir}) implies that this non-locality is mixing between UV in momentum and IR in position space
\cite{Ben-Israel:2015mda}. Hence, it is hard to see how interactions in the $b^{\dag}(p)$ modes basis can be summed in any useful form.
Therefore, it seems unlikely, in our opinion, that there is a stringy analog of  (\ref{yh}), but we cannot rule out this possibility.
This seems to leave us with two options:

1. There is no regular state at the horizon.

2. There is a regular state at the horizon, but we do not know how to construct it.

\section{Discussion}

About forty years ago, Israel showed that eternal black holes are regular also quantum mechanically \cite{Israel:1976ur}. Being so fresh and so clean, his argument appears robust. Yet, we argued that in string theory, at least for weakly coupled   strings in the $SL(2, \mathbb{R})_k/U(1)$ CFT background, the HH state is singular at the horizon. We are not able to rule out the possibility that there is a different state that is regular at the horizon. If such a state exists, the subtleties that go into writing it down might shed light on the firewall paradox. Needless to say that if it does not exist, this provides strong support to the firewall scenario not only in the collapsing case, but even for eternal black holes
(for which the arguments in \cite{Itzhaki:1996jt,Polchinski:2015cea,Marolf:2013dba,Braunstein:2009my,Mathur:2009hf,Almheiri:2012rt} do not apply).

It is natural to wonder if our discussion is valid only for the $SL(2, \mathbb{R})_k/U(1)$ black hole or if it is more general. Our reasoning is based on the fact that the reflection coefficient is known exactly in the $SL(2, \mathbb{R})_k/U(1)$ model. However, the origin of the unusual features of the stringy reflection coefficient is due to the condensate of the wound tachyon mode on the cigar, and this is expected
\cite{Kutasov:2005rr,Giveon:2012kp,Giveon:2014hfa,Mertens:2014dia,Mertens:2014saa}
to be rather general,
since a wound string is the order parameter associated with the fact that the thermal cycle is contractible at the tip. Hence,
even though we do not have an exact CFT description of, say, the Schwarzschild black hole, it is reasonable to suspect that similar phenomena take place there. If similar phenomena also take place in $AdS$ backgrounds, then reconsideration of the basic assumption in \cite{Maldacena:2001kr} -- that the horizon is smooth -- is called for.\footnote{The dual CFT evidence for the smoothness of the horizon (e.g. \cite{Kraus:2002iv,Fidkowski:2003nf,Hartman:2013qma}) is, in our opinion, too indirect to rule out this possibility. Incidentally, \cite{Barbon:2003aq} can be viewed as a CFT evidence for a non-smooth horizon.}

The relation between the wound string condensate and the singular HH state discussed here is likely to be more direct. We expect the condensate of the wound tachyon to be concentrated at a stringy distance from the tip (for large $k$). Hence, it likely affects the HH state in the way presented in figure \ref{winding}, namely, the singularity at the horizon discussed above is stretched to a stringy distance.
Presumably, the way to see this smearing in our construction is to include in the HH state the contributions of the stringy tower of massive modes.

\section*{Acknowledgments}

We thank Micha Berkooz, David Kutasov and Eliezer Rabinovici for discussions.
This work  is supported in part by the I-CORE Program of the Planning and Budgeting Committee and the Israel Science Foundation (Center No. 1937/12), and by a center of excellence supported by the Israel Science Foundation (grant number 1989/14).

\begin{appendix}
\section{Divergence in total energy and singular terms in the stress tensor}
In this appendix, we present detailed expressions concerning two points used in the paper: the divergence of the Minkowski energy of a HH-like state with a non-trivial phase, and the concentration of this divergent energy flux on the horizon. We perform all calculations in Rindler space with acceleration $a$, which in the small curvature limit is a good approximation for the black hole. In general, the Bogoliubov matrices $\alpha_{\omega k}$ and $\beta_{\omega k}$ are not identical in both cases, but for high frequencies (which would be interesting to us) the difference is negligible.

\begin{figure}
\centerline{\includegraphics[width=0.9\textwidth,clip=true,trim=145 180 130 180]{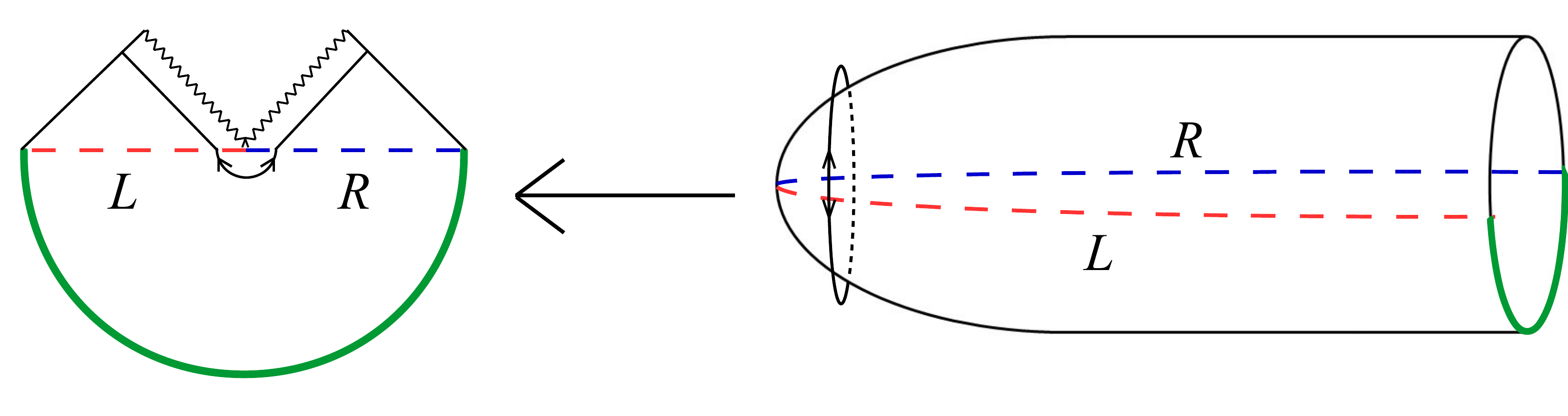}}
\caption{A winding string condensate around the tip of the cigar geometry would create  string excitations localized at a (presumably) stringy distance from the horizon in the Lorentzian geometry. The singularity in the physical black hole would be smeared over this stretched horizon.}
\label{winding}
\end{figure}

\subsection{The divergent Minkowski energy}

We define the phased state,
\begin{equation}
\ket{HH} = \frac{1}{\sqrt{Z}} \, \exp \left(\int d\omega \, e^{-\frac{\pi \omega}{a} +i \theta(\omega) } a_{\omega}^{R \dag} a_{\omega}^{L \dag} \right) \ket{0,0}~,
\end{equation}
where $a_{\omega}^{R/L}$ are the annihilation operators of a free massless scalar field in each Rindler wedge, $\ket{0,0}$ is the Rindler vacuum, and $\theta(\omega)$ is some non-trivial function.
When $\theta=0$, this state is the Minkowski vacuum.

We wish to inspect the Minkowski energy in the phased state $\ket{HH}$.
{}For this purpose, we first define a convenient operator basis for the Fock space. The annihilation operators in Minkowski space, denoted $b_k$, are related to the Rindler basis by a Bogoliubov transformation
(see e.g. \cite{Crispino:2007} for a review), so that
\begin{equation}
b_k = \int_0 ^\infty d\omega \left( \alpha_{\omega k}^R \, a_\omega^R + \left(\beta_{\omega k}^R\right)^* a_{\omega}^{R \dag} +\alpha_{\omega k}^L \, a_\omega^L + \left(\beta_{\omega k}^L\right)^* a_{\omega}^{L \dag}  \right)~, \label{bog}
\end{equation}
where $\alpha_{\omega k}$ and $\beta_{\omega k}$ are the Bogoliubov matrices. We can define an additional convenient basis,
\begin{equation}
B_\omega^{(1)} \equiv \cosh(x) \, a_\omega^L - \sinh(x) \, a_\omega^{R \dag}~, ~~~~~~~~~~ \tanh(x) \equiv e^{-\frac{\pi \omega}{a}}~, \label{Hmode}
\end{equation}
and a similar operator $B_{\omega}^{(2)}$, by taking $L\leftrightarrow R$ in (\ref{Hmode}). Here, $a$ is the Unruh acceleration, which in the  $SL(2, \mathbb{R})_k/U(1)$ model takes the value $a\to 1/\sqrt{2k}$.  Inverting (\ref{Hmode}) and plugging it into (\ref{bog}), we get
\begin{equation}
b_k = \int_0^\infty d\omega \sqrt{1-e^{-\frac{2\pi \omega}{a}}} \left( \alpha_{\omega k}^L \, B_\omega^{(1)} + \alpha_{\omega k}^R \, B_{\omega}^{(2)}  \right)~. \label{bog2}
\end{equation}
The dependence on $\alpha_{\omega k}$ alone is expected, since $B_\omega$ annihilates the Minkowski vacuum and, therefore, should not contain any creation operators.

\begin{figure}
\centerline{\includegraphics[scale=0.45]{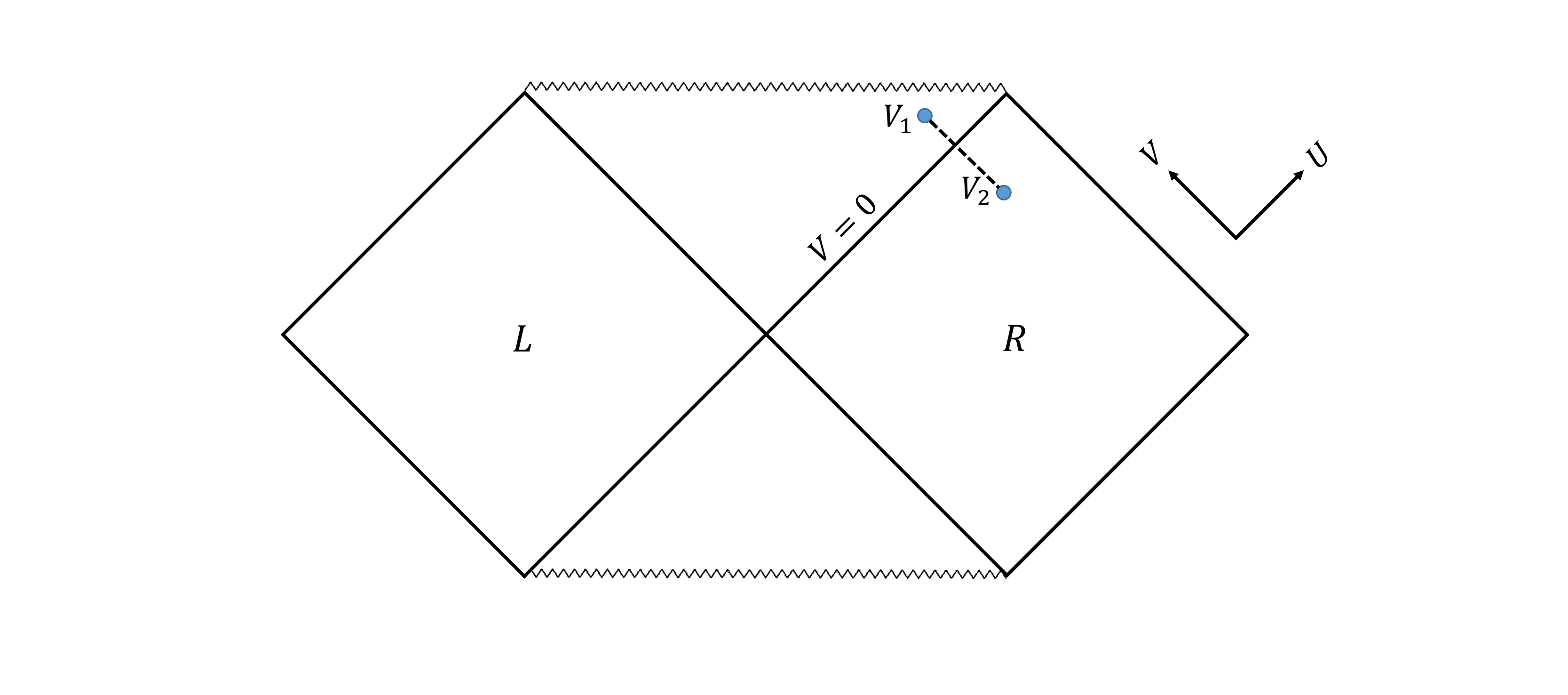}}
\caption{Point-splitting the stress tensor and evaluating $\langle \p \phi(V_1) \, \p \phi(V_2) \rangle$ shows that the only contributions to the Hamiltonian which cannot be taken to 0 by standard normal ordering are those for which $V_1 V_2 <0$, i.e., the split points are on opposite sides of the horizon. This forces the singular terms in $T_{VV}$ which lead to the divergence of the energy to be located exactly on the horizon.}
\label{Vcoord}
\end{figure}

We are interested in the total energy of the system. Since we are dealing with a free scalar field in flat space, the Minkowski Hamiltonian can be written as
\begin{equation}
H_{Min} = \int_0^\infty dk ~ k \, b_k^\dag b_k~,
\end{equation}
where we have explicitly taken the vacuum energy to zero. Plugging (\ref{bog2}) gives
\ben
H_{Min} &=& \int_0^\infty dk \, d\omega_1 \, d\omega_2 \,k \sqrt{ \left( 1-e^{-\frac{2\pi \omega_1}{a}}\right) \left( 1-e^{-\frac{2\pi \omega_2}{a}}\right)} \bigg( (\alpha_{\omega_1 k}^L)^* \alpha_{\omega_2 k}^L \, B_{\omega_1}^{(1)\dag} B_{\omega_2}^{(1)} \nonumber \\
&+&(\alpha_{\omega_1 k}^L)^* \alpha_{\omega_2 k}^R \, B_{\omega_1}^{(1)\dag} B_{\omega_2}^{(2)}  + (\alpha_{\omega_1 k}^R)^* \alpha_{\omega_2 k}^L \, B_{\omega_1}^{(2)\dag} B_{\omega_2}^{(1)} + (\alpha_{\omega_1 k}^R)^* \alpha_{\omega_2 k}^R \, B_{\omega_1}^{(2)\dag} B_{\omega_2}^{(2)}\bigg). \nonumber \\ \label{Emin}
\een
It is straightforward to calculate the expectation values in the $B_\omega$ basis; one gets
\begin{equation}
\langle B_{\omega_1}^{(1) \dag} B_{\omega_2}^{(1)}\rangle = \langle B_{\omega_1}^{(2) \dag} B_{\omega_2}^{(2)}\rangle = \left( \frac{\sin \left(\frac{\theta(\omega_1)}{2} \right)}{\sinh \left( \frac{\pi \omega_1}{a}\right)} \right)^2  \, \delta(\omega_1-\omega_2)~,
\end{equation}
and a vanishing result for the cross terms that appear in (\ref{Emin}). This means that
the Minkowski energy is
\ben
\langle H_{Min} \rangle
&=& \int_0^\infty dk \, d\omega \, k  \left( 1-e^{-\frac{2\pi \omega}{a}}\right) \, \frac{\sin^2 \left(\frac{\theta(\omega)}{2} \right)}{\sinh^2 \left( \frac{\pi \omega}{a}\right)} \left( \left|\alpha_{\omega k}^R \right|^2 + \left|\alpha_{\omega k}^L \right|^2   \right)~,
\een
and plugging (see \cite{Crispino:2007})
\begin{equation}
\alpha_{\omega k}^R = (\alpha_{\omega k}^L)^* = \frac{i e^{\pi \omega/2a}}{2\pi \sqrt{\omega k}} \left( \frac{a}{k}\right)^{-i \omega / a} \Gamma(1-i \omega/a)~,
\end{equation}
gives an occupation number for a Minkowski mode with energy $k$
\begin{equation}
\langle N_k \rangle \equiv \langle b_k^\dagger b_k \rangle = \frac{1}{\pi a} \, \frac{1}{k} \, \int_0^\infty d\omega \, \frac{\sin^2 \left(\frac{\theta(\omega)}{2} \right)}{\sinh^2 \left( \frac{\pi \omega}{a}\right)}~,
\end{equation}
and total energy
\begin{equation}
\langle H_{Min} \rangle = \frac{1}{\pi a}\int_0^\infty d\omega\, \frac{\sin^2 \left(\frac{\theta(\omega)}{2} \right)}{\sinh^2 \left( \frac{\pi \omega}{a}\right)}  \, \int_0^\infty dk~. \label{toteng}
\end{equation}
As expected, the result (\ref{toteng}) is vanishing for a trivial phase, but UV divergent for any non-trivial phase.

\subsection{Point-splitting the stress tensor}
In addition to the fact that since the horizon is the only orbit of the boost Hamiltonian on which putting a particle won't change the thermal behavior of the state $\ket{HH}$, there is another way to see the localization of the energy flux on the horizon: via explicit calculation of the stress tensor. We analyze its divergent terms through point-splitting, i.e, looking at the correlator $\langle \partial \phi(V_1) \, \partial \phi(V_2) \rangle$
of a free massless scalar field, where $\partial \equiv \partial_V$ is the left-moving derivative. In the limit $V_2 \to V_1$, this correlator gives the component $T_{VV}(V_1)$ of the stress tensor.

We begin by calculating some expectation values of the Rindler creation and annihilation operators,
\begin{eqnarray}
\langle HH | a^{R \dag}_\omega a^R_{\omega '} |HH \rangle =& \langle HH | a^{L \dag}_\omega a^L_{\omega '} |HH \rangle &= (e^{\frac{2\pi \omega}{a}} -1)^{-1} \, \delta(\omega-\omega '), \nonumber \\
\langle HH | a^R_\omega a^{R\dag}_{\omega '} |HH \rangle =& \langle HH | a^L_\omega a^{L\dag}_{\omega '} |HH \rangle &= (1-e^{-\frac{2\pi \omega}{a}})^{-1} \, \delta(\omega-\omega '), \label{expec} \\
\langle HH | a^R_\omega a^{L}_{\omega '} |HH \rangle =& ~~\langle HH | a^{R \dag}_\omega a^{L \dag}_{\omega '} |HH \rangle^* &= e^{i \theta (\omega)} (e^{\frac{\pi \omega}{a}}-e^{-\frac{\pi \omega}{a}})^{-1} \, \delta(\omega-\omega '), \nonumber
\end{eqnarray}
with all other combinations vanishing. The left-moving field $\Phi(V)$ can be expressed as (see \cite{Crispino:2007})
\begin{equation}
\Phi(V) = \int_0 ^\infty d\omega \left( a^R_\omega \, g^R_\omega(V) + a^{R \dag}_\omega \, g^{R \,*}_\omega(V) + a^L_\omega \, g^L_\omega(V) + a^{L \dag}_\omega \, g^{L \,*}_\omega(V) \right), \label{Phi}
\end{equation}
with
\begin{equation}
g^R_\omega (V)= \int_0 ^\infty \frac{dk}{\sqrt{4\pi k}} \left( \alpha^R_{\omega k} \, e^{-ikV} + \beta^R_{\omega k} \, e^{ikV} \right), \label{eigenf}
\end{equation}
and a similar expression for $g^L_\omega(V)$. Using (\ref{expec}),(\ref{Phi}),(\ref{eigenf}) and performing the $k$ integrations explicitly,
we get a result which depends on the relative sign of $V_1$ and $V_2$, i.e., whether both points are on the same side or different sides of the horizon (see figure \ref{Vcoord}).
{}For $\text{sign}(V_1) =\text{sign}(V_2)$, we get
\begin{eqnarray}
\langle \partial \phi (V_1) \, \partial \phi (V_2) \rangle &=& \frac{1}{4 a^2  \, \pi \, V_1 V_2} \int_0 ^\infty d\omega  \, \frac{\omega ~\cosh \left( \frac{\omega}{a} \middle( \pi -i \log \middle(\frac{V_1}{V_2} \middle) \middle) \right)}{\sinh \left(\frac{\pi \omega}{a} \right)}~, \label{sameside}
\end{eqnarray}
and for $\text{sign}(V_1) \neq \text{sign}(V_2)$, we take without loss of generality $V_1<0<V_2$, and get
\begin{eqnarray}
\langle \partial \phi (V_1) \, \partial \phi (V_2) \rangle &=& \frac{1}{4 a^2  \, \pi \, V_1 V_2} \int_0 ^\infty d\omega  \, \frac{\omega ~\cos \left( \theta(\omega) + \omega \tau \right)}{\sinh \left(\frac{\pi \omega}{a} \right)}~, \label{FNS}
\end{eqnarray}
with $\tau \equiv a^{-1} \log(-V_1/V_2)$. The result (\ref{sameside}) -- and in the $\theta=0$ case, (\ref{FNS}) as well -- can be regularized to give the expected result, $-\frac{1}{4\pi} \, \frac{1}{(V_1-V_2)^2}$.
These results are discussed in section 4.

\end{appendix}

\end{document}